# Enhancing Performance in Bolt Torque Tightening Using a Connected Torque Wrench and Augmented Reality

FAU Adeline[1,*], GHOBRIAL Mina[1,2,a], SEITIER Philippe[1,b], LAGARRIGUE Pierre[2,c], GALAUP Michel[3,d], DAIDIE Alain[1,e], and GILLES Patrick[1,f]

[1]Institut Clément Ader (ICA), Université de Toulouse, CNRS, UPS, INSA, ISAE-SUPAERO, IMT Mines Albi, 3 Rue Caroline Aigle, 31400 Toulouse, France

[2]ICA, Institut National Universitaire Champollion, Pl. de Verdun, 81000 Albi, France

[3]EFTS, Institut National Universitaire Champollion, Pl. de Verdun, 81000 Albi, France

*adeline.fau@insa-toulouse.fr, [a]ghobrial@insa-toulouse.fr, [b]seitier@insa-toulouse.fr, [c]pierre.lagarrigue@univ-jfc.fr, [d]michel.galaup@univ-jfc.fr, [e]daidie@insa-toulouse.fr, [f]gilles@insa-toulouse.fr

**Abstract.** Modern production rates and the increasing complexity of mechanical systems require efficient and effective manufacturing and assembly processes. The transition to Industry 4.0, supported by the deployment of innovative tools such as Augmented Reality (AR), equips the industry to tackle future challenges. Among critical processes, the assembly and tightening of bolted joints stand out due to their significant safety and economic implications across various industrial sectors. This study proposes an innovative tightening method designed to enhance the reliability of bolted assembly tightening through the use of Augmented Reality and connected tools. A 6-degrees-of-freedom (6-DoF) tracked connected torque wrench assists the operator during tightening, ensuring each screw is tightened to the correct torque. The effectiveness of this method is compared with the conventional tightening method using paper instructions. Participants in the study carried out tightening sequences on two simple parts with multiple screws. The study evaluates the impact of the proposed method on task performance and its acceptability to operators. The tracked connected torque wrench provides considerable assistance to the operators, including wrench control and automatic generation of tightening reports. The results suggest that the AR-based method has the potential to ensure reliable torque tightening of bolted joints.

## Introduction

Immersive technologies are increasingly used due to advancements in the performance of the available tools. Gandolfi provides a definition for the different immersive technologies [1]. Virtual Reality (VR) is the most immersive technology, as it creates a fully virtual environment. VR is particularly useful for tasks such as designing mechanical assemblies or training operators. Augmented Reality (AR) is less immersive and overlays digital information such as instructions or guidance onto the real world. Depending on the application, AR can be deployed using various devices, including remote screens, tablets, projectors or Head-Mounted Displays (HMDs). AR is particularly well-suited for training and educational purposes. Ghobrial *et al.* compared three instructional media for assisting with a machining center tool-loading procedure: paper-based instructions, video tutorials, and an AR scenario involving mechanical engineering students [2]. The results show that while the AR scenario has significant potential for instructional support, video-based instructions are more ergonomic, considering the hardware limitations, the specific tools used in the study and the given use case.

In the industrial sector, AR is primarily applied in the assembly, maintenance, and inspection phases to assist operators. Assembly, in particular, is a critical operation in the product manufacturing process. In their literature review, Wang *et al.* propose a chronological approach to the assembly process — before assembly, on-site assembly, and after assembly — to distinguish between Extended Reality (XR) applications [3]. Real-time guidance using AR presents an opportunity to assist operators during assembly operations. AR scenarios are used to guide operators through assembly tasks. The parts of a mechanical system are sequentially highlighted using virtual elements, based on a predefined assembly sequence or instructional order. The selection of AR content is crucial when creating scenarios, as it ensures clear communication to guide the operators through the task to be performed. Li *et al.* propose a classification of 2D and 3D visual elements for assembly processes [4].

For structural joints, bolted assemblies tightening is a critical safety issue. The aim for industrials is to minimize human errors during tightening, as torque tightening is often applied manually. One way to achieve this is through the use of Industry 4.0 technologies, in particular Augmented Reality which remains rarely used for torque tightening today.

The acceptance of AR in industrial applications remains a significant challenge despite its potential to enhance productivity, reduce errors, and support employee training. Quandt and Freitag highlight that user acceptance is a critical factor influencing the adoption of AR systems in maintenance, assembly, and training tasks [5]. While factors such as perceived usefulness and ease of use are essential, challenges persist, including ergonomic limitations of AR hardware, usability concerns, and cognitive load on users. Moreover, the integration of gamification, which could enhance engagement and motivation, is underexplored. Addressing these challenges through user-centered design and comprehensive field testing is crucial for widespread adoption.

Operators play a pivotal role in ensuring quality and efficiency in assembly, especially under High-Variety, Low-Volume (HVLV) conditions where product configurations and parts are constantly changing. According to Claeys *et al.*, poorly designed or incomplete instructions disproportionately raise operators' cognitive load, forcing them to guess or solve problems mid-process and eroding their trust in guidelines [6]. Meanwhile, Su *et al.* demonstrate that design and process complexities, such as challenging part shapes, mating directions, and part instability, can be a major source of operator mistakes, accounting for up to 20% of defects in a copier assembly case study [7]. Together, these findings highlight that both the clarity and completeness of instructions, as well as the minimization of product complexity factors, are crucial to reducing error rates and improving assembly outcomes. However, this poses a significant challenge in today's fast-paced, highly dynamic industries, where increasing product complexity and rapid innovation cycles often demand trade-offs between simplicity and functionality.

Device localization is fundamental to AR, allowing virtual content to be accurately anchored in the physical world. Modern AR devices, such as Head-Mounted Displays, rely on Visual-Inertial Odometry (VIO) combined with additional sensors like LiDAR to provide real-time 6 degrees-of-freedom (6-DoF) tracking. However, achieving accurate localization remains challenging, especially in large-scale, dynamic, or low-texture environments, where VIO can drift or fail to maintain consistent accuracy [8]. In addition to device localization, accurate 6-DoF tracking of objects is key in certain AR applications, particularly in industrial contexts where precise alignment of virtual content with physical tools and parts is essential. Methods for 6-DoF object tracking often rely on feature-based matching or edge-based tracking. Feature-based tracking methods use distinct points, textures, or shapes, while edge-based methods align object contours with detected scene edges. Each of the given technique handle low-texture or occluded objects more effectively. Combining these methods can enhance robustness, particularly for objects with diverse surface characteristics or challenging environments [9].

This study compares a new AR-based torque tightening method with paper-based tightening instructions, representing a conventional approach. It focuses on the assembly task performed by operators. The experiment evaluates the effectiveness and the acceptability of the AR guide with a 6-DoF tracked connected torque wrench. This new tightening method ensures that each screw has been tightened to the correct torque in the correct sequence. Execution time and the error rate in locating the tightened screws are analysed. These different metrics provide an overview of the performance of the two tightening methods.

**Experimental setup**

To evaluate the new tightening method proposed in this paper, the experimental approach involves operators tightening bolted assemblies. Participants follow instructions by performing a tightening sequence and applying a predefined torque to specific screws. During the study, each participant performs two tightening methods:
  (i)  a conventional tightening method using paper instructions for the tightening sequence;
  (ii) a tightening method using an AR guide and a 6-DoF tracked wrench.

Half of the participants completed the method with paper instructions followed by AR, while the other half started with the AR method and then the conventional method. Audio and video recordings of the scene are made while participants perform the two tightening methods.

The test environment consists of two parts to be assembled, and an additional part for participants to learn how to use the connected torque wrench and the AR assistant (Fig. 1a). One part is a grid with 50 tapped holes and the other is a flange with 13 tapped holes. The screws were all pre-installed so as not to favour any particular tightening sequence. For both tightening methods, participants begin by the tightening of the grid followed by the flange. The Wi-Fi connected torque wrench, a DynaSAM®4.0, supports tightening torques ranging from 1 to 10 N.m. A LED bar helps the operator during tightening by indicating the torque being applied (Fig. 2c). When the set torque is reached, the operator is alerted through a sound, wrench vibration, and the illumination of a red LED bar.

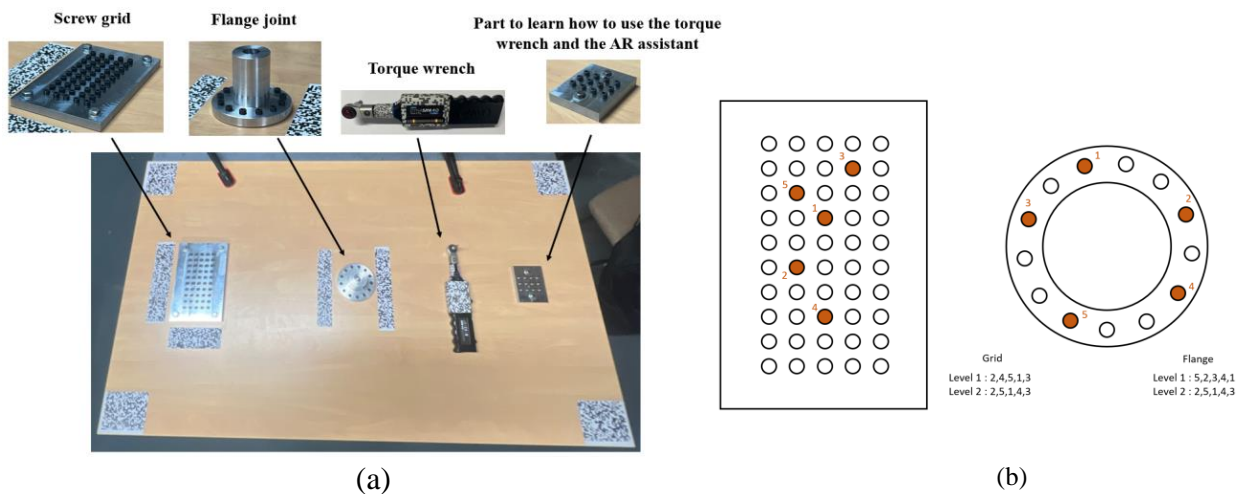

*Figure 1: (a) Assembly of mechanical parts during the experiment and the torque wrench used and (b) tightening sequence n°1 for the grid part and the flange showing the two torque levels applied on the screws*

The screws used are M5x8 in steel (8.8) screws with cylindrical hexagon socket heads. Each participant tightens 5 screws on the two parts following a predefined sequence (e.g., sequence n°1 in Fig. 1b). Two torque levels are required (3 N.m and 5 N.m). A total of 20 tightening operations is carried out by operators for each method. Each participant randomly selects one of three tightening sequences to be used for the two tightening methods.

Participants begin the experiment by completing a questionnaire to determine their profile (e.g., age, experience with Augmented Reality, use of a torque wrench, etc.). After each tightening method, participants complete two standard questionnaires about task load with the NASA Task Load Index (NASA-TLX) [10] and acceptability with the System Usability Scale (SUS) [11] to assess their perception of the method just tested.

- Participants

34 members of the research laboratory staff, aged between 20 and 60 years, participate in this study, including PhD students, professors, engineers and technicians. They are familiar with mechanical assemblies and have prior experience tightening screws. This condition was chosen to ensure a population representative of the knowledge level of assembly line operators in an industrial environment. Participants were excluded if they had previously performed these tightening sequences. 12% of the participants reported never using a torque wrench, and only 12% had experience with a connected torque wrench. 35% had never used XR technology before. Each tightening method was preceded by a period of acculturation and learning. This training involved performing several tightening operations on the part provided for this purpose (Fig. 1a). This ensured that the participants had a basic operational understanding of the tool before starting the experiment.

During the experiment, supervisors verify the position of the screws tightened by the participant. In addition, post-processing of the experiment for each participant confirms the position and torque of each tightening for the two methods by watching the video recordings. The recordings were also used to determine the execution time of the tightening operations for each method.

- Conventional tightening method

For the conventional tightening method, paper-based instructions are used (Fig. 2a). They describe the tightening sequences, the torque value to be applied and include a diagram showing the position of the screws to be tightened. A smartphone with the SAM Dyna® application defines the tightening method (torque, angle, etc.) and the torque to be applied. The maximum torque value achieved is stored in the application. At the end of the experiment, torque values are checked by the supervisors.

The operator controls the wrench from the application and manually sets the tightening parameters (set torque). Traceability of tightening operations for quality control is ensured by the operator, who fills in a document summarising the tightening operations carried out and the torque values applied to each screw.

The procedure for this method is as follows: the operator reads the instructions for the tightening of the grid part, sets the wrench parameters, tightens the grid part, reads the instructions for the tightening of the flange, sets the wrench parameters, and tightens the flange.

- Augmented Reality and 6-DoF tracked tool method

The same torque wrench is used as for the conventional tightening method but without the smartphone. In this method, an AR scenario is defined and combined with a 6-DoF tracked connected torque wrench. The Unity development platform was used to develop the Augmented Reality guide. This tightening method includes the Head-Mounted Display (Magic Leap 2), to keep the user's hands free, and the torque wrench (Fig. 2d). In Unity, a digital twin of the setup (table, parts, and wrench) was created, integrating torque behaviour for both bolts and the torque wrench. A custom script was developed (based on DynaSAM's documentation) to enable Wi-Fi communication between the Unity Build and the DynaSAM®4.0 torque wrench, the same way it communicates with the smartphone SAM Dyna® app to send torque settings and receive applied torque values in real time. The tightening method includes several specific features:

(i) Virtual elements, such as arrows, are used to indicate the position of the screw to be tightened and are superimposed on the real environment (Fig. 2c). The tightening sequence is defined in the scenario and the operator is required to follow it as instructed. The target torque and the applied torque values are displayed next to the screw head. Once the screws are tightened, a green cylinder is superimposed on the screw, to suggest to the operator that the screw is tightened at least to the minimum required torque (Fig. 2b and Fig. 2c).

(ii) The position of the wrench in relation to the system to be assembled can be determined using real-time tracking, using Vuforia Engine: A Model Target is defined for the main assembly (table with the three parts), which is static. This model target is initialized once by the supervisors. An advanced Model Target is used for the torque wrench, allowing automatic re-detection whenever the wrench's tracking is lost, without requiring the user to manually align the wrench with an initial view. This is necessary because the wrench is moved dynamically during tightening and must be tracked constantly. Due to the lack of detail when closing in on each part with the HMD, patterns were added at the corners of the table, and around the parts to minimize drift. A pattern was applied to the torque wrench's visible surface as well to improve tracking by enhancing visual key features. Given that the torque wrench is predominantly black, its surface lacks sufficient visual details, particularly during roll rotation, which can lead to tracking loss, especially given the dynamic nature of the task. Depending on the position of the wrench and the screw to be tightened, the torque setting is sent directly to the wrench. As a result, this eliminates the need for the operator to manually enter the torque setting. To enable the application to recognize the torque wrench as positioned on a specific bolt, a validation criterion was established: the screw bits of the torque wrench had to be within a specified distance from the bolt's head, a distance determined during initial trials prior to experimentation. Once this condition was met, the tightening of the target screw was validated when the applied torque reached the programmed value. This criterion was implemented to account for tracking inaccuracies.

(iii) The application kept track of the applied torque for each screw throughout the process, and at the end of a scenario, a summary report of the torques applied to each screw was automatically generated and saved locally in the headset, to be retrieved afterward.

This process consists of following the instructions given by the scenario. The user wears the Head-Mounted Display. Once the scenario has been started, the operator tightens the grid part and then the flange.

| Conventional tightening method | AR-based tightening method |
|---|---|

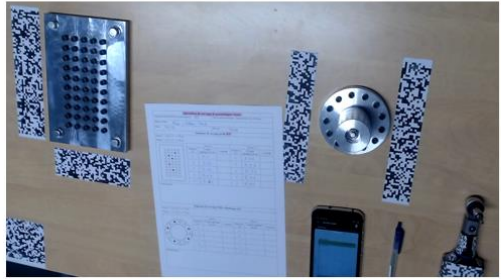

(a)

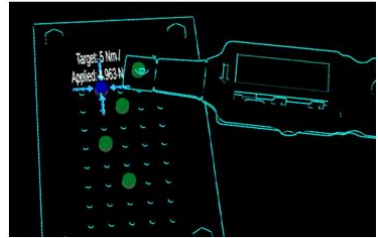

(b)

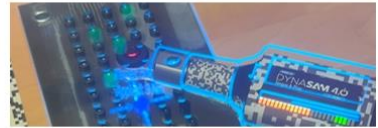

(c)

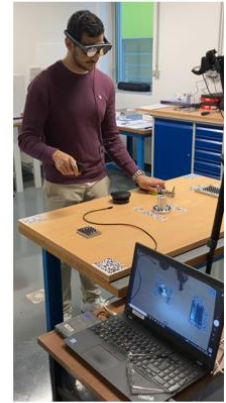

(d)

*Figure 2: (a) Paper tightening instructions and smartphone for the conventional tightening method, for the AR-based method (b) tracking of the grid part and the torque wrench, (c) virtual elements in the visual field of the user and (d) operator with the Head-Mounted Display*

**Results**

Several criteria are considered to analyse the results of the experimental approach:
  (i)  The user's perception of the tightening method;
  (ii) The operator's tightening performance.

These metrics cannot be separated in order to evaluate the new tightening method proposed as a whole. The tightening method includes the entire process. For the conventional tightening method, it corresponds to the tightening paper instructions, the application on the smartphone and the torque wrench. Only the Head-Mounted Display and the torque wrench are considered for the Augmented Reality and 6-DoF tracked tool method.

- User's perception of the tightening method

The acceptability of each tightening method was evaluated using the System Usability Scale (SUS) [11]. It consists of ten items, with positive and negative connotations, on a scale of 1 to 5, where 1 is the lowest usability and 5 is the highest. A subjective assessment of the tightening methods is therefore made by the operators. Participants indicated that they needed the assistance of a technician more often with the AR guide than with the conventional tightening method. This should be seen in light of the fact that 35% of the participants said that they had never used any XR technology before. For other items, such as complexity, ease of use, number of inconsistencies, self-confidence, etc.), the scores for the conventional tightening method and the AR-based method were very similar.

The NASA Task Load Index [10] measures operators' perceived workload by operators assessing factors related to the tightening method and workload. The scale starts at 1, indicating no workload and goes up to 20 for high workload. The NASA-TLX gives access to concepts such as mental demand, physical demand, time demand, effort, performance and frustration level. The task load review revealed approximately a 27% reduction in workload when using the AR-based method compared to the conventional tightening method. The AR-based method recorded a significantly lower perceived task load (5.1) compared to the conventional method (7.0), indicating that the operators found the AR system less demanding. This reduction in workload could be attributed to the AR interface's ability to simplify task execution by providing real-time visual guidance and automated adjustments. However, both methods had low NASA-TLX scores indicating a low workload. The reduction in workload with the AR-based method should be evaluated alongside user performance.

- Operator's tightening performance

User's performance is assessed based on two criteria: the time taken to complete the tightening task and the error rate in the aim of assessing productivity.

The time taken to complete the tightening tasks is related to efficiency and demonstrated a clear advantage for the AR-based method. The mean execution time was 5 minutes and 39 seconds divided into 2 minutes and 47 seconds for the grid and 2 minutes and 41 seconds for the flange. The mean execution time includes the transition time between the grid and the flange. For the conventional tightening method, the total average time was significantly longer reaching 10 minutes and 23 seconds (4 minutes and 54 seconds for the grid and 4 minutes and 26 seconds for the flange). Programming the wrench and reading the tightening instructions took the rest of the time. The faster execution with the AR-assisted method highlights its efficiency in guiding the operator, reducing the time needed for reading and interpreting paper-based instructions, as well as for side tasks like writing down the torque applied to each screw after each tightening operation. The error rate is defined by errors in locating the screw to be tightened, such as not following the defined sequence or selecting the wrong screw. Errors in setting the torque value in the application are also taken into account. The error analysis revealed a substantial reduction in errors when using the AR-based tightening method compared to the conventional approach. No operator-related errors were recorded in sequence order or screw localization. The AR system-imposed tightening sequence and clear positional guidance eliminated these common mistakes. It is important to note that some errors occurred due to application crashes or significant repositioning of the headset, which were caused by the low maturity of the prototype rather than user performance. Consequently, all the data collected from these instances were excluded from the analysis. 67.6% of participants made errors with the conventional tightening method. Users shifted their attention between tightening, reading the instructions and filling in the document, leading to numerous errors. 47% of participants made mistakes by following an incorrect tightening order, 29% misidentified screw locations, and 9% of the participants forgot to update the torque value programming during a required torque level transition. Error rates were higher during the grid tightening compared to the flange, likely because participants performed the grid first and sometimes corrected mistakes when moving to the flange. It is important to note that the total number of errors was not counted. Each type of error was recorded only once per participant, even if the participant made the same error multiple times.

- Synthesis of results

Table 1 shows the results of the 4 metrics for the two tightening methods. Similarly to Ghobrial *et al.* [2] the SUS scores are converted to a 0-100 scale. The results show a quasi-similar usability between the two methods with an average score of 73.1% for the conventional tightening method and 74.4% for the AR-based guide, 100% corresponding to the best usability.

*Table 1: Global results for the two tightening methods*

|  | **Conventional method** | **AR-based method** |
|---|---|---|
| Usability (SUS) | 73.1% | 74.4% |
| Task Load (NASA-TLX) (note out of /20) | 7.0 | 5.1 |
| Execution Time [seconds] | 623 | 339 |
| Number of people who made a mistake | 23 | 0 |

In Fig. 3 results are presented graphically with a radar representation. NASA-TLX scores are inverted and converted to a 0-100 scale for easier interpretation [2].

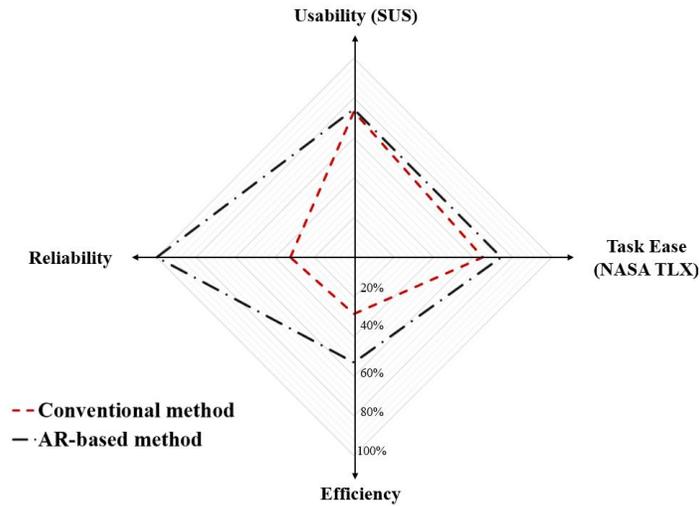

*Figure 3: Tool performance radar representation for the two tightening methods*

Operator efficiency is linked to the time taken to complete the tightening operations. It is calculated by dividing the minimum execution time for both methods by the mean execution time. With the AR-based method, none of the participants made any mistakes so the reliability is equal to 100%. Only 32.4% of the participants made no mistakes with the conventional tightening method using paper instructions. Fig. 3 shows that the AR method has a real impact on the operator's performance and the reliability of the tightening operations. The virtual guidance provided by the AR-based method resolved issues of sequence and screw localization, demonstrating its potential to improve the reliability of assembly operations.

**Discussion**

The study revealed several insights regarding operator behaviour and the impact of each method on performance. Although both methods are considered to be low workload and have similar SUS scores, the AR-based method is clearly superior in terms of user performance. Execution time is significantly reduced as the number of tasks to be carried out by the operator is greatly reduced. Conventional tightening required operators to manually document torque values, leading to frequent errors and increased workload. The AR-based method automated this process, ensuring complete traceability and eliminating documentation errors.

While both methods used the same torque wrench, participants showed more rigorous attention to the torque values applied with the conventional method. This discrepancy may result from:
  (i) Instances of wrench tracking issues in the AR-assisted method, causing operators to reapply torque, sometimes leading to over-tightening as a reflex.
  (ii) The lack of immediate torque feedback displayed in Augmented Reality, reducing operators' precision focus.

The current telecommunication limits impede global performance, requiring users to divide their attention between two separate torque level indicators, which is suboptimal. To address these challenges, a primary focus should be on ensuring real-time, rapid communication between the headset and the torque wrench. Displaying the applied torque values for previously tightened bolts could help operators develop a stronger association between the target torque and their motor skills.

The results highlight the general advantages of the AR-assisted method and its effectiveness. AR significantly accelerated the tightening process, with time savings that could be further enhanced as the technology matures. Sequence and localization errors were eradicated, emphasizing the effectiveness of AR in guiding relatively complex bolted assembly tasks. However, participants expressed concerns about the tracking reliability of the torque wrench. Improved software and hardware could enhance user experience and performance. In addition, the extended use of Head-Mounted Displays proved uncomfortable for some participants, highlighting the need for alternative AR display methods, such as projector-based systems. In our setup, the Magic Leap 2 headset was selected because it addresses many of the ergonomic challenges and hardware limitations highlighted by Ghobrial *et al.* such as weight and discomfort [2]. Moreover, as the application automates most interactions (e.g., torque setting, scenario navigation), users were not required to interact with the virtual interface, thereby mitigating further ergonomic concerns. Nonetheless, the method is equally transposable to other devices (e.g., the Hololens 2), provided similar compatibility with the Vuforia Engine and Unity development environment.

**Conclusion**

To improve assembly tightening reliability and reduce human error in the context of increasing production rates, new technologies are essential. Connected tools and Augmented Reality offer a viable one solution. For bolted assemblies, the aim is to improve the reliability of tightening operations by assisting the operator. A new tightening method is proposed. This method is based on Augmented Reality and a 6-DoF tracked connected torque wrench. It was compared with a conventional tightening method based on paper instructions. To achieve this, a test bench was designed with two parts on which operators performed torque tightening operations. Performance was gauged using metrics such as task load, acceptability and operator performance (execution time of the tightening task and error rate). The results showed that the proposed new method has the potential to assist the operator during tightening operations. The AR-based method can improve assembly task performance by reducing execution time and eliminating errors. User's perception is affected by the current level of technological maturity, with tracking difficulties, despite recognizing the method's potential. Wrench tracking and control needs to be improved to make the AR-based method easier to use, reducing latency-related difficulties. It will then be more accurate and robust. The use of Head-Mounted Displays, which are challenging for operators to wear for extended periods in industrial environments, may be questionable. Projecting virtual elements onto the parts to be assembled could offer a more acceptable solution. Achieving the same level of accuracy for the applied torque with the AR-based method as with the conventional method requires refinement in the choice of virtual elements. The maximum torque value applied to each screw can be specified. The encouraging results of this study carried out in a laboratory environment need to be confirmed in an industrial environment. In addition, more complex assemblies with bolts in several planes need to be considered. With an improved wrench tracking and relevant virtual elements, this bolt torque tightening solution should ensure that each screw is tightened to the correct torque and in the correct sequence.

**References**
[1] E. Gandolfi, Virtual Reality and Augmented Reality, Handbook of Research on K-12 Online and Blended Learning, 2nd ed., ETC Press, pp. 545–561, 2018.
[2] M. Ghobrial, P. Seitier, P. Lagarrigue, M. Galaup, and P. Gilles, Effectiveness of machining equipment user guides: A comparative study of augmented reality and traditional media, presented at the Material Forming. (2024) 2320-2328. https://doi.org/10.21741/9781644903131-255.

[3] B. Wang, L. Zheng, Y. Wang, W. Fang, and L. Wang, Towards the industry 5.0 frontier: Review and prospect of XR in product assembly, Journal of Manufacturing Systems. 74 (2024) 777-811. https://doi.org/10.1016/j.jmsy.2024.05.002.

[4] W. Li, J. Wang, S. Jiao, M. Wang, and S. Li, Research on the visual elements of augmented reality assembly processes, Virtual Reality & Intelligent Hardware. 1 (2019) 622-634. https://doi.org/10.1016/j.vrih.2019.09.006.

[5] M. Quandt and M. Freitag, A Systematic Review of User Acceptance in Industrial Augmented Reality, Front. Educ. 6 (2021) 700760. https://doi.org/3389/feduc.2021.700760.

[6] A. Claeys, S. Hoedt, E.-H. Aghezzaf, and J. Cottyn, Assessing assembly instructions quality using operator behavior, Int J Adv Manuf Technol. 135 (2024) 4531-4548. https://doi.org/10.1007/s00170-024-14740-z.

[7] Q. Su, L. Liu, and D. E. Whitney, A Systematic Study of the Prediction Model for Operator-Induced Assembly Defects Based on Assembly Complexity Factors, IEEE Trans. Syst., Man, Cybern. A. 40 (2010) 107-120. https://doi.org/10.1109/TSMCA.2009.2033030.

[8] P.-E. Sarlin *et al.*, LaMAR: Benchmarking Localization and Mapping for Augmented Reality, in Computer Vision – ECCV 2022, S. Avidan, G. Brostow, M. Cissé, G. M. Farinella, and T. Hassner, Eds., in Lecture Notes in Computer Science, Cham: Springer Nature Switzerland. 13667 (2022) 686-704. https://doi.org/10.1007/978-3-031-20071-7_40.

[9] T. Hodaň *et al.*, "BOP: Benchmark for 6D Object Pose Estimation," in Computer Vision – ECCV 2018, V. Ferrari, M. Hebert, C. Sminchisescu, and Y. Weiss, Eds., in Lecture Notes in Computer Science, Cham: Springer International Publishing. 11214 (2018) 19-35. https://doi.org/10.1007/978-3-030-01249-6_2.

[10] S. G. Hart and L. E. Staveland, Development of NASA-TLX (Task Load Index): Results of Empirical and Theoretical Research, Advances in Psychology. Elsevier. (1988) 139-183. https://doi.org/10.1016/S0166-4115(08)62386-9.

[11] A. Bangor, P. Kortum, and J. Miller, Determining What Individual SUS Scores Mean: Adding an Adjective Rating Scale, J. Usability Studies. 4 (2009) 114-123.